\documentclass[12pt]{article}
\usepackage[cp1251]{inputenc}
\usepackage{amsmath}
\usepackage{amssymb}
\usepackage{cite}
\usepackage[russian]{babel}
\usepackage{graphicx}
\Rus \textheight 24 true cm \textwidth 16.5 true cm \hoffset -10true mm \voffset -25 true mm

\begin{document}
\large
\renewcommand{\abstractname}{}
\renewcommand{\refname}{References.}
\author{Gusev A.V., Rudenko V.N., Yudin I.S.}
\title{
\begin{flushleft}
{\small 519.246; 524}
\end{flushleft}
Low frequency signals of large scale GW-interferometers
\date{ Lomonosov Moscow State University, \\
Sternberg Astronomical Institute, \\
Moscow, Russia }}

\maketitle
\author
\begin{abstract}

\noindent Application of the large scale gravitational wave interferometers for  measurement of geophysical signals at very low frequencies is considered. Analysis is concentrated on the mechanism of penetration of quasistatic geophysical perturbation through the main interferometer output. It is shown that it has a parametrical nature resulted in slow variations of the optical transfer function of the interferometer. Geophysical modulation index is calculated for any harmonical component of the output spectrum, but mainly for a photon circulation frequency appeared in the case of stochastic illumination of modes neighbour to the central resonance. Value of the effect is estimated for different operational regimes of the device. For improvement of geophysical signal readout a modernization of the instrument with using of two component resonance optical pump is proposed and a correspondent calculation is carried out. Numerical estimations for different regimes of the setup are given together with discussion of possible application for measuring some weak gravitational effects.           


\underline{\textbf{key words:}} large scale gravitational wave
interferometers, gravitational geodynamical effects, precise
optical measurement technique.\\
e-mail:rvn@sai.msu.ru \\
\end{abstract}

\textbf{1. Introduction.} \\
A possibility of applicaion of large scale gravity wave interferometers
(LIGO, VIRGO) for a registration geophysical effects at very low frequencies was discussed in the number of papers \cite{1,2,3,4,5,6,7} It was supposed that a geophysical information as a by-product can be read out from feed back circuits controling coordinate and angular position of mass-mirrors for to keep the operational tuning of the interferometer. This tuning corresponds to the complete distractive interference at the main interferometer output, so called a "dark spot"\, regime. Obviously voltages of correcting drivers are proportional to geophysical deformations of the interferometer base. First experimental realization of these ideas was demonstrated  at  the  VIRGO  interferometer \cite{8} during scientific runs VSR-1,VSR-2. It was proved that the large scale gravitational wave interferometer can  successfully operate as a very sensitive two coordinate strain meter. Idea of using it as an angular gravity gradiometer \cite{2} still did not find a confirmation at practice because a nontrivial complexity of angular correction circuits in the VIRGO setup.\\
At the same time it was shown at LIGO interferometers that tidal geophysical perturbations can be registered also at the main relatively high frequency
 output. Such perturbations appeared in some indirect way through the ampitude modulation a free spectral range harmonic, or photon circulation frequency, in the arm FP cavities \cite{9,10,11}. Attempt to explain this effect by relativistic gravilational variation of the light velocity in the tidal gravitational background (under condition of fixed mirrors) was failed: the observable scale of modulation was at three order of value larger the relativistic effect forecast \cite{11,12}. Later traces of quasi tidal modulation ("siderial periodicity") of other harmonical components at the main output were detected also in the VIRGO setup. It was occured in the process of searching for gravitational wave signals from Vela pulsar at frequencies on the order of $20\,Hz$ \cite{13}. In the paper of authors \cite{14} for a saving the hypothesis of relativistic nature of modulation effect a supposition of "non precise tuning" \, in the dark spot of destructive interference was accepted, so called the regime of operation in a "gray spot" \, \cite{14} Formally with a tiny selection of "detuning"\, from the condition of "dark spot"\, one could fit the observable value of modulation effect with the theoretically expected one. However in operational practic of LIGO and VIRGO setups such "tiny detuning" \, specially never was applied.\\
In such situation an explanation of the modulation effect might be given with a refusing from the hypothesis of "position fixed mirrors"\, at least at the level of accuracy of position correcting drivers. Such level for the VIRGO interferometer consists $10^{-12} m$ (or $10^{-15}$ in the term of arm deformation) that is just on three-four orders larger  the estimate of the relativistic gravitational variation of optical length.

Objective of this paper consists in a more detailed analysis of the mechanism of penetration of slow deformation signals through the main interferometer output.
At first we calculate a transfer function of the interferometer following to a simlified optical scheme (OS) presented at the $fig.1$\, where the recycling mirror is omitted. Optical pump is taken in the form typical for a phase modulated light. Then the transfer function is studied at special frequency bands: zones of the main optical resonance and its neighbour modes as well as for sidebands of radio frequency modulation. Synchronous photo detection of the light flux  at modulation frequency produces an output current containig in particular the photon circulation frequency which can be separated with a proper band pass filter. It is shown that the amplitude of this frequency is modulated by slow geophysical variations of the interferometer arms. Analysis of final formulae allows to explain the observation of tidal perturbations during S5 LIGO serie. \\
Further research comes back to the model of monochromatic pump  when the photon circulation frequency does not appear at the main output. It is shown that even in this case a geophysical perturbations can be read out at the main output factually through the amplitude modulation of any its spectral harmonics.
It is occured due to the parametrical convertion of slow variation of arms into pump sideband components with the following demodulation through the photodetection process. Probably by such mechanism one could partly explaine a detection of the durnal tidal harmonics in the VIRGO output data \cite{13}. \\
Analysis of the mechanism of registering interferometer base slow variations allows to formulate a setup modernization (introduction of two component pump) for to readout geophysical signals in a more robust manner. At the end we discuss the influence of recycling and possible application of low frequency readout for a measurement of weak geophysical effects and for a detection of very low frequency gravitational waves. \\

\textbf{2. Setup model.} \\

The figure 1. shows a principle layout of the LIGO and VIRGO
interferometers. The Michelson configuration is complicated by the
presence of multi reflective Fabry-Perot cavities in arms,which are described below as equivalent complex mirrors. To simlify calculations in this paper we omitted an additional so called "recycling mirror" $\,$
between the laser and beam splitter. It does not affect results of
subsequent analysis at least qualitatively. To keep the arm
mirrors at the optical resonance by control circuits a phase modulation of
the input beam at the radio frequency ($\sim 25 MHz$ for LIGO) is introduced (EOM and RF units at the fig 1.).  Synchronous photodetection of the FP-reflected light at the modulation frequency (in the frame of the Pound-Drever scheme \cite{15}) allows to form the driver voltages for correction of mirror positions. Informative signal of theinterferometer as a gravitational wave detector appears at the destructive interference output (the main or antisymmetric port)\, in the form of low frequency photocurrent components. Using a narrow band filter one can in principle select a desirable part of signal spectrum. In particular in the papers \cite{9,10,11} a similar filter was performed as a synchronous detection at the fre spectral range frequency.

After such remarks we present below a formal description of beams interaction in the optical scheme (OS) fig.1. using the method of "complex envelope" (see appendix A1). \\
At the interferometer input there is a phase modulated optical pump with the carrier frequency $\omega_{0}$
$$
E_{1}(t)=\mathrm{Re}\,\left[\tilde{E}_{1}(t)\exp\left\{j\omega_{0}t\right\}\right],
\;
\tilde{E}_{1}(t)=\tilde{E_{0}}(t)\exp\left\{j\beta\sin\bar{\Omega}t\right\},
\eqno(1)
$$
so that the complex envelope $\tilde{E}_{1}(t)$ contains harmonics
of the modulation frequency $\bar{\Omega}$ with a modulation index  $\beta$. In general the amplitude envelope $\tilde{E}_{0}$ may be described as a large constant component $A$ (line amplitude) at a small stochastic
background (line pedestal) $\tilde{a}(t)$
$$
\tilde{E}_{0}(t)=A + \tilde{a}(t)
$$
It is supposed that the complex process $\tilde{a}(t)$  is changing
much slow the period of radio modulation. It means a spectral width of
the pedestal is much less the modulation frequency but it might be
larger the free spectral range of arm cavities. For small
modulation indexes ($\beta<1$) the exponent expansion in (1) leads
to
$$
\tilde{E}_{1}(t)\simeq(A +
\tilde{a}(t))\left[J_{0}(\beta)+J_{1}(\beta)\left(
\exp\left\{j\bar{\Omega}t\right\}-\exp\left\{-j\bar{\Omega}t\right\}\right)\right].\eqno(2)
$$
where $J_{k}(\cdot)$ is the Bessel function on the order of $k$.\\
Aim of our subsequent consideration is a getting output signals in the scheme fig.1 taking into account slow geophysical perturbations of the interferometer base. For a comparison with experiment we will calculate a so called "geophysical modulation index" \, as a ratio of the output voltage variable part to its permanent level. This universal parameter is conenient because it does not depend on particular details of the setup like the pump intensity, depth of Pound-Drever modulation, photodetection efficiency ect. But for estimation the "signal to noise" \, ratio a knowledge of such parameters  of course will be required.\\ 

\textbf{3. Interferometer response with the phase modulated pump.}

Electromagnetic field $E_{2}$ at the output of the interferometer
OS also can be presented as a quasi harmonic process:
$$
E_{2}(t)=\mathrm{Re}\,\left[\tilde{E}_{2}(t)\exp\left\{j\omega_{0}t\right\}\right]
$$
Let us introduce a pulse reaction $g(t)$ and transfer function
$G(\omega)$  of the OS
$g(t)=\mathrm{Re}\,\left[\tilde{g}(t)\exp\left\{j\omega_{0}t\right\}\right]\leftrightarrow G(\omega)$.\\
Then the output field $E_{2}$ is defined by the Dugamel integral equation (see appendix A2)
$$
\begin{array}{c}
E_{2}(t)=E_{1}(t)\ast
g(t)\rightarrow\tilde{E}_{2}(t)=\tilde{E}_{1}(t)\ast\tilde{g}(t), \\
\tilde{g}(t)\leftrightarrow\tilde{G}(\omega),\;\tilde{G}(\omega)=2G(\omega_{0}+\omega),\;|\omega|\ll\omega_{0}.
\end{array}\eqno(3)
$$
Then using (2) and (3) one comes to
$$
\begin{array}{c}
\tilde{E}_{2}(t)=  (1/2)(A + \tilde{a}(t)
)(\left\{\frac{}{}J_{0}(\beta)\tilde{G}(0)+J_{1}(\beta)
\left[\tilde{G}^{(-)}(\bar{\Omega})\cos\bar{\Omega}t +
j\tilde{G}^{(+)}(\bar{\Omega})\sin\bar{\Omega}t\right]\right\}, \\
\tilde{G}^{(+)}(\bar{\Omega})=\tilde{G}(\bar{\Omega})+\tilde{G}(-\bar{\Omega}),\;
\tilde{G}^{(-)}(\bar{\Omega})=\tilde{G}(\bar{\Omega})-\tilde{G}(-\bar{\Omega}).
\end{array}\eqno(4)
$$
At the photodetector output the field $E_{2}$  produces oscillation at modulation frequency with voltage
$v(t)\propto|\tilde{E}_{2}(t)|^{2}$ presented by slow quadratures
$$
v(t)= v_{c}(t)\cos\bar{\Omega}t - v_{s}(t)\sin\bar{\Omega}t
$$
After the Pound-Drever phase detection the quadrature components will appear as
$$
\begin{array}{c}
v_{s}(t)\propto\mathrm{Re}\,\left[\left(A\tilde{G}(0)+\tilde{a}(t)\ast\tilde{g}(t)\right)^{*}j
\tilde{G}^{(+)}(\bar{\Omega})\right], \\
v_{c}(t)\propto\mathrm{Re}\,\left[\left(A\tilde{G}(0)+\tilde{a}(t)\ast\tilde{g}(t)\right)^{*}
\tilde{G}^{(-)}(\bar{\Omega})\right]. 
\end{array}\eqno(5)
$$
To get a structure of the output signal in a more clear form one needs to
calculate the OS transfer function spectral densities involved in
formulae (5). \\

\textbf{3.Transfer function of the OS with compound mirrors.} \\

Transfer function of optical configuration presented at the fig.1 corresponds to the Michelson interferometer with compound mirrors. Role of such mirrors belongs to arm Fabry-Pero cavities.
Let $R_{1}(\omega)$ and $R_{2}(\omega)$ are reflection coefficients of these mirrors. Supposing the absolute (hundred percents) reflection for the arm end mirrors one has a simple formula for FP reflectivity \cite{12}
$R_{i}(\omega)=   \exp\left\{j\phi_{i}(\omega)\right\},\;i=1,2.$
Here $\phi_{i}(\cdot)$ is the phase difference $\phi_{i}(\cdot)$  between an insident optical wave and one reflected from the compound arm mirror.
It depends on a detuning between the pump frequency and optical resonance
frequency of FP cavity. \\
Simple calculation for the optical configuraion at fig.1 results in
$$
G(\omega)=\frac{1}{2}\exp\left\{-j\omega\tau\right\}R_{2}(\omega)
\left[1-\exp\left\{2j\omega\frac{\Delta
l}{c}\right\}\frac{R_{1}(\omega)}{R_{2}(\omega)}\right] \eqno(6a)
$$
or in a more compact form it is written as
$$
G(\omega)=\exp\left\{-j\omega\tau\right\}R_{2}(\omega)
H(\omega).\eqno(6b)
$$
Between  three factors in the formula (6b) a principal role
belongs to the third one - $H(\omega)$. Fist two factors do
not change the modulus of transfer function and may introduce only
some inessential delay time in the response. In our calculation
below we will use $H(\omega)$ as enough good approximation of
$G(\omega)$ , i.e.
$$
\tilde{H}(\omega)=2H(\omega_{0}+\omega)\simeq1 -
\exp\left\{2j(\omega_{0} +\omega)\frac{\Delta l
}{c}\right\}\exp\left\{j\psi(\omega)\right\}, \eqno(7)
$$
here
$\psi(\omega)=\varphi_{1}(\omega_{0}+\omega)-\varphi_{1}(\omega_{0}+\omega)$
presents a difference of two arm cavitiy resonance phase shifts
$\varphi_{i}(\omega)=\
arg R_{i}(\omega)$, $i=1,2$. \\
Our special interest belongs to the transfer function at
resonance regions of FP-cavities as well as at radio sideband
frequencies. \\

\textbf{3.1 OS transfer function at the arm resonance zones.} \\

Let's note as $\Omega_{1,2}(k)$ resonance frequencies of both
arms in a k-th resonance zone which are shifted in respect of the
main (central) resonance $\Omega(0)$ according to the relation
$$
\Omega(k)=\Omega(0)+k2\pi\nu,\;k=0,\pm1,\ldots,
$$
here $\nu=c/2L$ is the free spectral range frequency of the arm FP
cavity.
Its values for each arm are not equal, although their difference is very small $\nu_{1}\simeq\nu_{2}$.\\
Physical principle of GW interferometer consists in registration of small arm lengths changing induced by the incoming gravitational wave. For the FP cavity it means a correspondent variation of optical resonance frequencies. Below we
will take into account a shift of these frequencies under
arm deformation $\xi\rightarrow \xi_{1},\,\xi_{2}$ 
$$
\Omega(0)=\omega_{0}(1+\xi), \eqno(8)
$$
Using the formula (7) after simple transformation one comes
to the transfer function for a $k$- resonance zone
$$
\tilde{H}(\Omega(k))\simeq1-\exp\left\{j\left(\omega_{0}\frac{2\Delta
l}{c}+\psi_{0}\right)\right]\exp\left\{jk\frac{\pi}{1-r}\frac{\Delta
L }{L}\right\}, \eqno(9)
$$
This transfer function depends on both tuning factors: "arms
length" \, difference $\Delta L $ and "michelson length" \, difference
$\Delta l$.
 They are not independent in the process of "operation
point keeping" \,- a variation of one of them causes the
correspondent change of the second.\\
From the condition of "dark spot"\, $\tilde{H}(\Omega(0))=0$ at the main resonance one comes to the specific value of "michelson length difference "\, which we will call the "optimal"\, one $\Delta{l}=\Delta{l_{opt}}$
$$
\Delta l_{\mathrm{opt}}=\overline{\Delta l}-\lambda\frac{\psi_{0}}{4\pi},\eqno(10a)
$$
where the other special notations were introduced 

$$
\overline{\Delta l}=m\frac{\lambda}{2},\;\psi_{0}=\frac{2\pi}{(1-r)}\frac{L}{\lambda}\Delta\xi , \eqno(10b)
$$
thus the relative phase shift\, $\psi_{0}$\, in (9) depends on the large arms length  difference $\Delta\xi=\xi_{1}-\xi_{2}$. \\

In operational regime the control system keeps the arm lengths
constant at quasistatic frequencies (say below $1 Hz$) with the
definite accuracy $\Delta\xi_{st}\sim 10^{-15}$ (on the order of $10^{-12}m$
in absolute value for Virgo setup). Above these frequencies arms have to be considered as a free ones and GW signals can be registered as arm's deformation
$\Delta\xi_{gw}$. So the total deformation of arms can be presented as
a sum of residual quasistatic term (uncompensated by control system) and relatively fast dynamical perturbation in the GW detection frequency region ($(10 - 10^{4})Hz$ for Virgo setup).
$$
\Delta\xi=\Delta\xi_{st}+\Delta\xi_{gw},
$$
The correspondent relative phase shift $\psi_{0}$ also contains
these parts
$$
\psi_{0}=\psi_{0}^{(st)}+\psi_{0}^{(gw)}.
$$
$$
\psi_{0}^{(st)}=\frac{2\pi}{(1-r)}\frac{L}{\lambda}\Delta\xi^{(st)},\;
\psi_{0}^{(gw)}=\frac{2\pi}{(1-r)}\frac{L}{\lambda}\Delta\xi^{(gw)}.
$$
The transfer function in resonace zones (6b),\, (9) then can be read as
$$
\tilde{H}_{\omega,k}\simeq-j\left[\frac{4\pi}{\lambda}(\Delta l-\overline{\Delta l})+
\underbrace{\psi_{0}^{(st)}+\psi_{0}^{(gw)}}_{\psi_{0}}
+k\frac{\pi}{(1-r)}\frac{\Delta L}{L}\right], \eqno(11)
$$
or in the main resonace zone $k=0$
$$
\tilde{H}_{\omega,0}\simeq-j\left[\frac{4\pi}{\lambda}(\Delta l-\overline{\Delta l})+\psi_{0}^{(st)}+\psi_{0}^{(gw)}\right], \eqno(12)
$$
It is naturaly to suppose that in the process of "operation point keeping"\, only the quasistatical part of $\psi_{0}$ participates, so the condition of "dark spot"\,(10a) has to be read as
$$
\Delta l_{\mathrm{opt}}=\overline{\Delta l}-\frac{\lambda}{4\pi}\psi_{0}^{(st)}.
$$
At practice however there is always some small shift (error of precise tuning)
$\delta l$, so that  
$$
\Delta l=\Delta l_{\mathrm{opt}}+\delta l=\overline{\Delta l}-\frac{\lambda}{4\pi}\psi_{0}^{(st)}+\delta l , \eqno(13)
$$
then the formula (11) is transformed as 
$$
\tilde{H}_{\omega,k}\simeq-j\pi\left[\frac{4}{\lambda}\delta l+\frac{\psi_{0}^{(gw)}}{\pi}+k\frac{1}{(1-r)}\frac{\Delta l}{L}\right],\eqno(14)
$$
For estimation of the noise variance at the otput of scheme fig.1 it is useful to know a module of combined transfer function for two neighbour modes symmetrical to the main resonance. It looks like
$$
\left|\tilde{H}_{\omega,1}\right|^{2}+\left|   \tilde{H}_{\omega,-1}\right|^{2}\propto
\left[\frac{4}{\lambda}\delta l
+\frac{\psi_{0}^{(gw)}}{\pi}\right]^{2}+\left[\frac{1}{(1-r)}\frac{\Delta L }{L}\right]^{2}, \eqno(15)
$$
and we will use these formulae (14),(15) under a calculation low frequency signals at the output of the scheme fig.1.\\

\textbf{3.2\, OS transfer function at radio sideband zones. }\\

Besides the resonance zone it is important to know of the transfer function at  sideband frequencies which play a role of heterodyne waves under photodetection process in the "dark spot" regime. Coming back to the formulae (6),(7) and taking into account that at sideband frequencies reflection coefficients have a trivial form $R_{i}(\omega_{0} \pm\bar{\Omega})\simeq-1$,one finds the expression
$$
\tilde{H}(\omega\pm\bar{\Omega})\simeq1-\exp\left\{2j(\omega_{0}\pm\bar{\Omega})\frac{\Delta
l }{c}\right\}\simeq-j\left(-\psi_{0}^{(st)}+ 4\pi\frac{\delta
l}{\lambda}\pm2\bar{\Omega}\frac{\Delta l}{c}\right), \eqno(16)
$$
Calculation of the photodetector output (4),\,(5) requires combined
transfer functions of both symmetrical sidebands
$\tilde{H}^{(+)},\, \tilde{H}^{(-)}$ introduced in the formulae (4),\,(6b).
Using also the expression (14) one can get finally these transfer
functions as
$$
\left\{
\begin{array}{cc}
\displaystyle{
\tilde{H}_{\omega}^{(+)}(\bar{\Omega})=-2jQ_{c},\;Q_{c}\simeq\left(4\pi\frac{\delta l}{\lambda}-\psi_{0}^{(st)}\right);} \\
\displaystyle{
\tilde{H}_{\omega}^{(-)}(\bar{\Omega})=-2jQ_{s},\;Q_{s}\simeq2\overline{\Omega}\,\frac{\Delta l}{c}.}
\end{array}\right.\eqno(17a)
$$
where the functions $Q_{c}$,$Q_{s}$ are defined in details as
$$
\left\{
\begin{array}{cc}
{\displaystyle
Q_{c}\simeq\frac{2\pi L}{\lambda}\left(\frac{2\delta l }{L}-\frac{\Delta\xi_{st}}{(1-r)}\right);} \\
\displaystyle{
Q_{s}\simeq
\frac{2\pi L}{\lambda}\cdot\frac{2\bar{\Omega}}{\omega_{0}}
\left(\frac{\overline{\Delta l}}{L}-\frac{2\pi}{(1-r)}\Delta\xi_{st}+
\frac{\delta l}{L}\right).}
\end{array}\right.\eqno (17b)
$$
In our analysis below a special interest presents the folowing combination of quadrature components 
$$
Q_{c}^{2}+Q_{s}^{2}\propto\left(\frac{2\delta l}{L}-\frac{\Delta\xi_{st}}{(1-r)}\right)^{2}+
\mu^{2}\left(1-\frac{2\pi}{(1-r)}\frac{L\Delta\xi_{st}}{\overline{\Delta l}}+
\frac{\delta l}{\overline{\Delta l}}\right)^{2},\eqno(18a)
$$
where a small parameter $\mu$ is introduced as
$$
\mu=\frac{2\bar{\Omega}}{\omega_{0}}\frac{\overline{\Delta l}}{L}.\eqno(18b)
$$\\
It is clear from (18b) that $\mu \ll 1$. In the formula (18a) $\mu$ has the multiplier in round baskets. Below however this factor will be replaced by unit with good approximation.\\ 

\textbf{4. Noise variance at the region of the circulation frequency.} \\

Let's come back to the formulae of quadrature components (5) for to calculate 
correspondent noise quadratures at the output of optical system fig.1
Taking into account the spectral composition of amplitude $\tilde{E}_{0}(t)$ (the line and pedestal) one comes to spectral densities of the quadratures $v_{s}(t),\,v_{c}(t)$
$$
\begin{array}{cc}
\displaystyle  
{N_{c}(w) = N_{0}(\omega)\left[|H_{\omega}(\omega)|^{2}+|H_{\omega}(-\omega)|^{2}\right]Q_{c}^{2}}; \\
\displaystyle
{N_{s}(w)=N_{0}(\omega)\left[|H_{\omega}(\omega)|^{2}+|H_{\omega}(-\omega)|^{2}\right]Q_{s}^{2}.} 
\end{array}\eqno (19) 
$$
The mutual (cross) energetic spectrum $N_{c,s}(\omega)$ is described by the following expression 
$$
N_{c,s}(\omega)\propto jN_{0}(\omega)\left[\left|\tilde{H}_{\omega}(\omega)\right|^{2}-
\left|\tilde{H}_{\omega}(-\omega)\right|^{2}\right]Q_{c}Q_{s}
$$
Processes $v_{c}(t)$,\,$v_{s}(t)$ are  incorrelated ones in coincident moments of time. It means $ |H_{\omega}(\omega)|= |H_{\omega}(-\omega)|$ and $N_{c,s}(\omega) =0$. \\
For comparison with experiment we will use the output variance $\sigma_{0}^{2}$ after synchronouse detection at the free spectral range frequency (or arm circulation frequency) It can be calculated as a noise spectral density averaged into a frequency window $\Delta\omega$  
$$
\sigma_{0}^{2}=\frac{1}{\pi}\int\limits_{\nu-\Delta\omega}^{\nu+\Delta\omega}\left\{N_{c}(\omega)+
N_{s}(\omega)\right]d\omega\simeq\frac{2}{\pi}\left[N_{c}(\nu)+N_{s}(\nu)\right]\Delta\omega
$$
then from (19)one can find
$$
\sigma_{0}^{2}\propto\left(Q_{c}^{2}+Q_{s}^{2}\right)\left[\left|\tilde{H}_{\omega,1}\right|^{2}+
\left|\tilde{H}_{\omega,-1}\right|^{2}\right],\eqno(20)
$$
here $\Delta\omega$ is the filtering bandwidth for the complex stochastic process  $v_{0}(t)$. With (15), (18a), (19) the equation (20) might be written as 
$$
\displaystyle
{\sigma_{0}^{2}\propto \left[\left(\frac{2\delta l}{L}-\frac{\Delta\xi}{(1-r)}\right)^{2}+
\mu^{2}\right]} 
{\left[\left(\frac{4\delta l}{\lambda}+\frac{\psi_{0}^{(gw)}}{\pi}\right)^{2}
+\frac{1}{(1-r)^{2}}\left(\frac{\Delta L}{L}\right)^{2}\right]},\\
\eqno(21)
$$
This formula will be used in the resulting discussion below. \\

\textbf{5. Geophysical signal at the circulation frequency.} \\

Physical picture of optical fields interaction at the output 
photo detector consists in a nonlinear mixture of the light of neghbour symmetrical modes and pump sidebands playing role of heterodyne 
radiation. It is reflected in the formula (20) which presents the production 
squared transfer functions of modes and sidebands. Both depends on
quasistatic variations of large arms $\Delta\xi_{st}$. After coherent radio 
detection at the modulation frequency the circulation (or free spectral range) component will appear in the composition of low frequency output. A key condition for appearence of this component is the hypothesis
of "noise illumination"\, of modes neighbour to the central resonance.
In contrast with the paper \cite{12} (where only one neighbour mode was considered) we took into account both symmetrical modes. It produces
a serious difference in estimation of the geophysical signal value. To perform such estimation one has to analyse the output variance $\sigma_{0}^{2}$ presented by formula (21). However a result depends on the "operational point" tuning.\\
A) $\it {"dark \, spot" regime.}$ \\   
The full distructive imterference at the main output, or "dark spot" regime, was considered as a main operational regime for the GW laser interferometers. It is believed it corresponds to minimum optical noises. 
Taking $\delta l=0$ in the formula (21) one concludes that the term with slow geophysical variations $\Delta\xi_{st}$ exists in the output variance $\sigma_{0}^{2}$ only in the second order of value. In papers \cite{11,12} the estimate of variance contained the first order of $\Delta\xi_{st}$.  Such result was received due to an ignoring the fact of simultaneous luminosity of two neghbour modes around the central line. \\
However experimental data of LIGO also demonstrated the "linear effect"\, so that it contents the main tidal harmonics of Earth deformation.To overcome this contradiction one might to suppose some deflection from the "dark spot"\, regime. Analysis of the  tranfer function (21) shows  that the linear dependence of the output variance $\sigma_{0}^{2}$ from slow deformations $\Delta\xi_{st}$ is possible with refusing from idealistic "dark spot"\, regime and admitting some illuminosity  due to a small tuning shift $(\delta l\neq0)$ according to the formula (13). \\   
B) $\it {"grey \, spot" regime.}$ \\
There are two conceivable ways how one can come to the "grey spot" regime.\\
a) First model is similar to that used in the "dark spot" regime. Control circuits keep the operational point position which in this case corresponds to the tuning condition
$$ 
\Delta l=\Delta l_{\mathrm{opt}}+\delta l,\;\delta l=\mathrm{const}\neq0
\eqno(22)
$$
i.e. the constant shift $\delta {l}$ from the "dark spot" condition $\Delta l=\Delta l_{\mathrm{opt}}$ provides a finit lminosity in a resulting interference picture.\\
From (18a),(20) one can easy get
$$ 
\sigma_{0}^{2}\propto Q_{c}^{2}+Q_{s}^{2}\propto\left(\frac{2\delta l}{L}-\frac{\Delta\xi_{st}}{(1-r)}\right)^{2}+
\mu^{2}.\,
$$
This variance contains the permanent and variable parts, so 
the relative modulation index is estimated by the formula
$$
m'\simeq\frac{2\delta l}{L}\frac{\Delta \xi}{(1-r)}{\left[
\left(\frac{2\delta l}{L}\right)^{2}+\mu^{2}\right]^{-1}}. \eqno(23)
$$
b)\,Second model, reflects a situation with small geophysical perturbations when $\Delta\xi_{st}$ is less the controlling accuracy of feed back circuts. Then the compensation voltage in michelson arms is absent and their difference remains to be constant $\Delta l = const$. The formula (13) can be rewritten as
$$
\delta l=\Delta l-\overline{\Delta l} + \frac{\lambda}{4\pi}\psi_{0}^{(st)}= \delta l_{0} + \frac{\lambda}{4\pi}\psi_{0}^{(st)}, \eqno(24)
$$
where a new note of permanent detuning was used $\delta l_{0}= \Delta l-\overline{\Delta l}$. \\
In the same manner like above one can get using (15),(20),(21)
$$
\sigma_{0}^{2}\propto \left[\left|\tilde{H}_{\omega,1}\right|^{2}+
\left|\tilde{H}_{\omega,-1}\right|^{2}\right] \propto \left[\frac{4}{\lambda}\delta l\right]^{2}+\left[\frac{1}{(1-r)}\frac{\Delta L }{L}\right]^{2}, \eqno(25)
$$
Combining formulae (24),(25) one gets the modulation index
$$
m\simeq\frac{L\Delta\xi_{st}}{(1-r)\delta l_{0}}\left[1+\left(\frac{\lambda}{4(1-r)\delta l_{0}}\right)^{2}
\left(\frac{\Delta L}{L}\right)^{2}\right]^{-1}.\eqno(26)
$$
If one neglect $\mu^{2}$ in the formula (23) and the small correction of unit in quadratic baskets of formula (26) they both will practically coincide so as $ \delta l$ and $\delta l_{0}$ have the same sense of small detuning for shifting from the "dark spot"\, to "gray spot"\, regime. Thus the general estimate of modulation index is
$$
m\simeq\frac{L\Delta\xi_{st}}{(1-r)\delta l},\,\,\delta l\approx \delta l_{0} \eqno (27)
$$
It is easy to see that selecting the magnitude of "gray spot detuning" \, $\delta l$ one can change the modulation index.\\
Let us consider the interesting estimates.\\
i) \textit{Relativistic effect of optical refractive index variations in the tidal gravitational potential} \cite {12} \\
In this case $\Delta \xi \approx 10^{-19}$ ; the other parameters: $L=4\cdot 10^{5} cm, \,(1-r)\simeq 10^{-2}$. A substitution these data in (27) even with very small (unrealistic ) meaning of detuning $\delta l = 10^{-4} \lambda $ results in $ m\simeq 0.04\% $ on two orders of magnitute less the observable in experiment value \cite{9}.\\
ii)\textit{Effect of residual tidal deformation of the interferometer arms.}\\
The accuracy of mirror's position keeping consists $10^{-10}$ cm. If one supposes a presence of residual tidal deformation of large arms on this level then it has the order $\Delta\xi\simeq10^{-15}$. Under a resonable detuning $\delta l=10^{-2}\lambda$ one comes to the estimate of modulation index $m\simeq 4\% $ in a good agreement with obsevable effect \cite{9}.\\

\textbf{6. Geophysical signal at arbitrary frequencs.} \\

Above we have analysed the way of penetration a geophysical information  at the photon circulation (or FSR) frequency for to explaine  observations presented in reports \cite{9,10,11}. As we believe it is the result of slow variations OS transfer function. Residual arm deformation changes FP mode resonance frequencies and through the parametrical mechanism they are upcoverted to the optical range frequencies. After photodetection and coherent demodulation they can be readout through the corresponding low pass filtering.\\
Now we have to remark that photon circulation frequency is only one specific harmonic in the spectrum of output signal. Moreover it requires the special condition,- a luminiosity of modes neghbour to the central resonance. Meanwhile
it is clear that slow variations of OS transfer function have to affect also any other harmonics in the output spectrum. Really using our formalism it is easy to show the arbitrally choosed ouput harmonic will be amplitude modulated by geophysical perturbations.\\
Without requirement of having the FSR component at the main output one can consider the model with a pure harmonical pump at central resonance zone, i.e. $\tilde{E}_{0}(t)=A $ and  $\tilde{a}(t)=0 $. Then from the formula (5) for a quadature component, say $v_{c}(t)$, one has
$$
v_{c}(t)\propto  A\mathrm{Re}\,\left[\tilde{G}^{*}(0)\tilde{G}^{(-)}(\bar{\Omega})\right] \propto A\mathrm{Re}\,\left[\tilde{H}^{*}(0)\tilde{H}^{(-)}(\bar{\Omega})\right].
\eqno{(28)}
$$
Coming back to the operation in the dark spot $(\Delta l=\Delta l_{\mathrm{opt}} \,,\delta l=0)$ one can get from the formula (14)
$$
k=0,\delta l=0:\;\tilde{H}^{*}(0)\simeq j\psi_{0}^{(gw)}.
\eqno{(29)}
$$
(it is worth to remark that for quasistatic deformation $\Delta\xi^{(st)}$ the transfer function $\tilde{H}^{*}(0)$ equal to zero due to "dark spot"\,regime)
Under condition $\Delta l=\Delta l_{\mathrm{opt}}$ OS transfer function at sidebands \, $( frequencies \omega_{0}\pm\bar{\Omega})$\, (17.а) is read as
$$
\tilde{H}^{(-)}(\bar{\Omega})\simeq -2jQ_{s},Q_{s}\simeq2\bar{\Omega}\frac{\Delta l_{\mathrm{opt}}}
{c}. \eqno{(30)}
$$
Combining formulae (28),(29),(30) one comes to the following result
$$
v_{c}(t) \propto A\psi_{0}^{(gw)}\bar{\Omega}\frac{\Delta l_{\mathrm{opt}}}{c}\propto\psi_{0}^{(gw)}\left(1-\frac{1}{2(1-r)}\frac{L}{\overline{\Delta l}}\Delta\xi^{(st)}\right). \eqno{(31)}
$$
So as $\psi_{0}^{(gw)}$ presents some harmonic in the output signal spectrum of the GW interferometer one has to conclude looking at the formula (31) that the amplitude of such harmonics will be modulated by slow geophysical variations $
\Delta\xi^{(st)}$ with a modulation index given by
$$
m\simeq\frac{L\Delta\xi_{st}}{2(1-r)\overline{\Delta l}}. \eqno (32)
$$
Comparison (32) with the formula (27) shows that the parameter $\delta l\sim \delta l_{0}$ in (27) is replaced here by larger value $\overline{\Delta l}$ (10b). It looks as the "geophysical modulation effect"\, is decreased in respect the one for circulation frequency. However this reduction takes place in the "dark spot"\, regime. Meanwhile in such condition at circulation frequency the effect would be absent at all (the "gray spot"\, would be required).\\

\textbf{7. Setup with the two component pump.} \\

Above we have analised possible mechanisms of penetration of geophysical
signals into the main output of the GW interferometer. The effect might be large  at free spectral range frequency in the "gray spot"\, regime. However it looks like some stochastic parasitic phenomenon arised due to amplitude fluctuation of the optical pump. In order to make the effect more regular and reliable one could introduce a two component pump providing the simultaneous luminosity of two neighbour modes. In this case the appearance of enough power circulation frequency component at the main output will be guaranteed. Below we present briefly a calculation of the effect value for the two component pump. \\
Let's suppose that input laser pump $E_{0}(t)$ presented by superposition
$E_{0}(t)=E_{0,0}(t)+E_{0,1}$ of two narrow band oscillations 
$(E_{0,k}=\mathrm{Re}\,\tilde{E}_{0,k}(t)\exp\left\{j\omega_{i}t\right\})$
with close resonance frequencies $\omega_{0}\sim\omega_{1}=\omega_{0}+2\pi\nu$. 
Then for the complex amplitude $\tilde{E}_{0}(t)$ in formula (1) one has to put
$$
\tilde{E}_{0}(t)=\tilde{E}_{0,0}(t)+\tilde{E}_{0,1}(t)\exp\left\{j2\pi\nu t\right\}.\\
\eqno{(33)}
$$
where $\nu$ is the circulation (or FSR) frequency.
As in past complex envelopes $\tilde{E}_{1}(t)$ end $\tilde{E}_{2}(t)$
are coupled as 
$$
\begin{array}{cc}
\tilde{E}_{1}(t)=\tilde{E}_{0}(t)\left[J_{0}(\beta)+2jJ_{1}(\beta)
\cos\bar{\Omega}t\right], \\
\tilde{E}_{2}(t)=\tilde{E}_{1}(t)\ast\tilde{g}(t),
\end{array}
$$
In the quasistatic approximation ( $\tilde{E}_{0}(t)$ considered as a slow variable in respect of $\cos\bar{\Omega}t$ and $\sin\bar{\Omega}t$) one comes to  
$$
\tilde{E}_{2}(t)\simeq J_{0}(\beta)\left[\tilde{E}_{0}(t)\ast\tilde{g}(t)\right]
+2jJ_{1}(\beta)\tilde{E}_{0}(t)\left[\cos\bar{\Omega}t\ast\tilde{g}(t)\right].\\
\eqno{(34)}
$$
Now the correspondent analog of the formula (4) looks like 
$$
\tilde{E}_{2}(t)\simeq J_{0}(\beta)\left[\tilde{E}_{0}(t)\ast\tilde{G}(0)\right]+jJ_{1}(\beta)
\tilde{E}_{0}(t)\left[\tilde{G}^{(+)}(\bar{\Omega})\cos\bar{\Omega}t+
j\tilde{G}^{(-)}(\bar{\Omega})\sin\bar{\Omega}t\right],
$$
where the $\tilde{G}^{(+)}$ and $\tilde{G}^{(-)}$ are defined by (4).\\
At the photodetector output quadrature components of the circulation frequency 
are defined by the following expressions
$$
\begin{array}{cc}
v_{c}(t)\propto\mathrm{Re}\,\left[\left[\tilde{E}_{0}(t)\ast\tilde{g}(t)\right]^{*}j\tilde{E}_{0}(t)
\tilde{G}^{(+)}(\bar{\Omega})\right], \\
v_{s}(t)\propto\mathrm{Re}\,\left[\left[\tilde{E}_{0}(t)\ast\tilde{g}(t)\right]^{*}\tilde{E}_{0}(t)
\tilde{G}^{(-)}(\bar{\Omega})\right].
\end{array}
\eqno{(35)}
$$
Distinction of these expressions from (5) consists in the accounting of time dependence of the complex amplitude $\tilde{E}_{0}(t)$\,(33)  (the constant part $\tilde{E}_{0}$  was omitted in (5) ). So the formulae (35) \, take into account nonlinear terms of photodetection.\\
Let us consider a simple case of two component pump 
$$
\tilde{E}_{0,0}(t)=A_{0},\;
\tilde{E}_{0,1}(t)=A_{1}\exp\left\{j\vartheta\right\},\;A_{1}=\mathrm{const}.\\
$$
At practice the initial phase $\vartheta $ is likely unknown.
Below it considered as some stochastic variable homogeniously distributed at the interval $(0,2\pi)$.\\

Then(33) reduces to 
$$
\tilde{E}_{0}(t)=A_{0}+A_{1}\exp\left\{j(\nu t+\vartheta)\right\} \\
\eqno{(36)}
$$
then the first factor in (35) looks like
$$
\left[\tilde{E}_{0}(t)\ast\tilde{g}(t)\right]=A_{0}\tilde{H}(0)+A_{1}\exp\left\{j(\nu t+\vartheta)\right\}\tilde{H}(\nu).
$$
here according to (6b) we used $\tilde{H}$ for transfer function instead of full form $\tilde{G}$. \\
Filtering the output signal in a narrow band around the circulation frequency
results in 
$$
\begin{array}{cc}
\left[\tilde{E}_{0}(t)\ast\tilde{g}(t)\right]^{*}\tilde{E}_{0}(t)\simeq A_{0}A_{1}
\left[\left(H(0)+\tilde{H}(\nu)\right)^{*}\cos(\nu t+\vartheta)+ \right.\\
\left.j\left(\tilde{H}(0)-\tilde{H}(\nu)\right)^{*}\sin(\nu t+\vartheta)\right]+
\textit{high harmonics}
\end{array}
$$
For ransfer function at resonance modes $H(0)$ and $H(\nu)$  one can use the formula (11) taking $\psi_{0}$ expressed through the $\Delta{\xi}$. Then the first (resonance mode) factor in (35) is written as \\
$$
\tilde{H}(k\nu)\simeq\tilde{H}(0)-j\pi k\frac{1}{(1-r)}\frac{\Delta L}{L},\;
k=0,\pm1,\ldots,
$$
with 
$$
\tilde{H}(0)\simeq-j\frac{4\pi}{\lambda}\left[\left(\Delta l-\overline{\Delta l}\right)+
\frac{L}{2(1-r)}\Delta\xi_{st}\right].
$$
Introducing new notes $M^{(+)}$ and $M^{(-)}$, so that
$$
\begin{array}{cc}
\left(\tilde{H}(0)+\tilde{H}(\nu)\right)^{*}\simeq j\pi M^{(+)}, \\
\left(\tilde{H}(0)-\tilde{H}(\nu)\right)^{*}\simeq -j\pi M^{(-)},
\end{array}
$$
where
$$
\begin{array}{cc}
\displaystyle{
M^{(+)}\simeq\frac{8}{\lambda}
\left[\left(\Delta l-\overline{\Delta l}\right)+\frac{L}{2(1-r)}\Delta\xi_{st}+
\frac{\lambda}{8(1-r)}\frac{\Delta L}{L}\right],}\\
\displaystyle{M^{(-)}\simeq\frac{1}{(1-r)}\frac{\Delta L}{L}=\mathrm{const},}
\end{array}
\eqno{(38)}
$$
one presents the first (resonance mode) factor in the compact form

$$
\begin{array}{cc}
\left[\tilde{E}_{0}(t)\ast\tilde{g}(t)\right]^{*}\tilde{E}_{0}\simeq jM^{(+)}\cos(\nu t+\vartheta)+ \\
M^{(-)}\sin(\nu t+\vartheta)+\textit{high harmonics}.
\end{array}
\eqno{(39)}
$$
Transfer function at the sidebands $\left(\tilde{H}^{(+,-)}(\bar{\Omega})\right)$  from (16),\,(17a) can be written as
$$
\tilde{H}(\pm\bar{\Omega})\simeq-j\frac{4\pi}{\lambda}\left[\left(\Delta l-\overline{\Delta l}
\right)+\frac{\bar{\Omega}}{\omega_{0}}\Delta l\right].
$$
with their required combinations
$$
\left\{
\begin{array}{cc}
\displaystyle{
\tilde{H}^{(+)}(\bar{\Omega})\simeq jQ_{c},\;Q_{c}=-8\pi j\frac{\Delta l-\overline{\Delta l}}{\lambda},} \\
\displaystyle
{\tilde{H}^{(-)}(\bar{\Omega})\simeq jQ_{s},\;Q_{s}=8\left(\frac{\bar{\Omega}}{\omega_{0}}\right)
\frac{\left(\Delta l-\overline{\Delta l}\right)+\overline{\Delta l}}{\lambda}.} 
\end{array}\right.
\eqno{(40)}
$$
Substitution (39),\,(40) into (35) results in
$$
\begin{array}{cc}
v_{c}(t)\propto\mathrm{Re}\,\left[\left(jM^{(+)}\cos(\nu t+\vartheta)+M^{(-)}\sin(\nu t+\vartheta)\right)\times j(-2j)Q_{c}\right]+\ldots,\\
v_{s}(t)\propto\mathrm{Re}\,\left[\left(jM^{(+)}\cos(\nu t+\vartheta)+M^{(-)}\sin(\nu t+\vartheta)
\right)(-2j)Q_{s}\right]+\ldots\\
\end{array}
$$ 
and finally one has  
$$
\begin{array}{cc}
v_{c}(t)\propto Q_{c}M^{(-)}\sin(\nu t+\vartheta)+\ldots, \\
v_{s}(t)\propto Q_{s}M^{(+)}\cos(\nu t+\vartheta)+\ldots.
\end{array}
\eqno{(41)}
$$
It is likely that for unknown initial phase $\vartheta$ the optimal observable
variable (or sufficient statistics) will be a quadratic value of the envelope of oscillation at circulation frequency
$$
\Theta(\Delta\xi_{st})=\left(Q_{c}M^{(-)}\right)^{2}+\left(Q_{s}M^{(+)}\right)^{2}.
\eqno{(42)}
$$
Expanding this expression on power of small parameter $\Delta\xi_{st}$ one can estimate the amplitude modulation index as
$$
m=\left[\frac{d}{d\Delta\xi_{st}}\ln\Theta(\Delta\xi_{st})\right]_{\Delta\xi_{st}=0}\\
\eqno{(43)}
$$
To illustrate this formula let's consider a simple typical situation in which
$\Delta l=\overline{\Delta l}$ i.e. the "michelson arm difference" \, is equal to the integer number of optical half wave (in general it is some arbitrary "gray spot"\, regime).
Then from (38),\,(40) one collects 
$$
Q_{c}=0,\;Q_{s}=8\left(\frac{\bar{\Omega}}{\omega_{0}}\right)=\mathrm{const},
$$
$$
M^{(+)}=\frac{\Delta L}{(1-r)L}\left[1+\left(\frac{L\Delta\xi_{st}}{\lambda}\right)\frac{L}{\Delta L}\right],\;
M^{(-)}=\mathrm{const}.
$$
At last from (42),\,(43) one estimates the modulation index value as
$$
m\simeq\left(\frac{L\Delta\xi_{st}}{\lambda}\right)\frac{L}{\Delta L}.
\eqno{(44)}
$$
The all three formulae (27),\,(32),\,(44) for the "geophysical modulation effect"\, will be compared below.\\

\textbf{8. Results and discussion.} \\

The goal of our analysis consisted in a more clear understanding (compare with the papers \cite{11 , 12}) of the mechanizm of penetration very slow geophysical perturbations at the main relatively high frequency output of large GW interferometers. Now one can formulate the following  conclusions:\\
a).The geophysical signals produce an amplitude modulation of the photon circulation (or FSR ) frequency. This component appears at the main output due to some "parasitic"\, luminosity of modes neighbouring to the main resonance. So it is a consequence of nonideal monochromaticity of the interferometer light pump (let's call it as "real pump") containing beside the narrow central line also a small intensitive but enough frequency wide (larger the FSR interval) spectrum of amplitude fluctuation. The slow (geophysical) perturbations of the interferometer base affect its optical transfer function. Due to this the circulation frequency harmonic is occured to be amplitude modulated at geophysical low frequencies. In principle it allows to read out a geophysical information. Our more rigorous analysis (compare with \cite{11,12}) shows that the effect of "geophysical modulation" disappears in the first order due to compensative contributions of pair of neighbour FP modes symmetrical to the main resonance in the "dark spot"\, regime. It was shown that a reconstruction of the linear effect is possible under refusing from the "dark spot"\, regime and an admitting a some residual interference luminosity, i.e. a shifting to the "gray spot"\, regime. A more-less realistic tuning of the residual luminosity allows to explain through such mechanism the observation of tidal harmonics at LIGO setups resulting from the residual arms deformation beyond the sensitivity threshold of mirror's position control circuits. \\
b). With the monochromatic light pump (called as "ideal pump"\,) the photon circulation frequency is absent in the output signal spectrum. However also in this case the geophysical information might be read out from the main output. This time it can be done through the amplitude demodulation of arbitrary selected harmonic from the output spectrum. Reason of this consists in the same dependency of interferometer optical transfer function on a residual deformation of interferometer base.  But the amplitude modulation index of such selected harmonic will be much less then in the case with photon circulation frequency. At the same time the harmonic intensity might be high the intensity of neighbour modes illuminated by amplitude fluctuation of pump. Partly it has to compensate a decreasing of modulation index. \\
c). The understanding described in the points (a, b) stimulated a considering of concevable interferometer with the two component pump illuminating simultaneously two resonace FP modes ("double pump"\, model). In this model one come back to the manner of "geo measurement"\, through the circulation frequency but with some advantages. Now there is no "a compensation contribution of symmetrical mode"\, and intensity of the neighbour mode becomes equal to central resonance one (much high the amplitude fluctuation level). Such modernization leads to a large enhancement of the geophysical response value and crucially improves a "signal-to-noise"\, ratio at the intrinsic noise background. So a "geo sensitivity"\, of the setup will be defined only by environmental noises
(the case of ideal measuring device).  \\
 For a numerical illustration it is interesting to compare relative value of modulation indexis (27),(32),(44) using required parameters from VIRGO and LIGO setups. So as the mechanizm of geo-signals registration is associated with residual (incompensated by control circuits) part of arm variations, the value of quasistatic deformation $\Delta\xi_{st}$ has to be substituted as the $10^{-15}$ and less. Arm lengths are $ L=(3\,-\,4)$\,km. Assimetry of arms roughly has the orders: $\Delta L \sim 1\,cm$ and $\overline{\Delta l}\sim\, 30\,cm$ (see \cite{11}). Detuning parameter can be taken on the order of FP cavities resonance width $\delta l\approx 10^{-2} \lambda$ .\\
At first, let's compare the cases of "real"\, (27)\,$(m=m_{r}$( and "ideal"\, (32) $(m= m_{i})$ pumps. Its ratio is estimated as
$$
\frac{m_{r}}{m_{i}}=\frac{\overline{\Delta l}}{\delta l}\sim 10^{7}
$$
proving the fact of small effect for the one monochromatic pump. \\
At second, a comparison of cases of the "double" (44) $(m=m_{d})$ and "real"\,pumps resulted in 
$$
\frac{m_{d}}{m_{r}}=\frac{\delta l}{\Delta l}\frac{L}{\lambda}(1-r)\sim 1
$$
so modulation indexes for the "double"\, and "real"\, pumps are comparable.
Both correspond to operation in a "gray spot"\, regime.\\

Beside these estimates it would be desirable to evaluate the level of output
geo-signal which depends not only on the modulation index but also on amplitude of the proper carrier. In the case of "real"\, pump it is the amplitude of FSR harmonic at the photodiod output in the fig.1 \, It appears as a result of heterogine mixing of the neighbour mode harmonic and radio sideband component of the pump. For LIGO setups the neighbour mode harmonic had the spectral level $E_{\pm}\approx 10^{-7}E_{0}\,Hz^{-1/2}$ \cite{11}. In the mode resonance bandwidth $\sim 200\,Hz$ it resulted in the amplitude standard $ E_{\pm}\sim  \textasciitilde10^{-6}E_{0}$. \\
For the case of "ideal"\, pump one needs to know a fluctuation standard of output harmonics in a zone of maximum sensitivity. Roughly one can estimate it from the LIGO sensitivity curve using a simplest formula for the FP interferometer signal intensity response $\Delta I$ induced by its base variation: $\Delta I \approx I_{0}\,(2\pi L/\lambda)F (\Delta x/L)$. The deformation value is estimated
according to the relation $(\Delta x/L)\approx h_{\nu}\sqrt{\Delta\nu} \approx h_{min}\sim 10^{-21}$ (it was used $h_{\nu}\sim 10^{-22}Hz^{-1/2}, \,\,\Delta\nu\sim 200\,Hz$ for the zone of maximum sensitivity). Finally one comes to the standard of output harmonics value $<\Delta E_{fl}>\approx
\sqrt{\Delta I}\sim 10^{-3}E_{0}$ much larger the stochastic amplitude of the neighbour modes. \\
Now one can estimate a relative value of geo-signals $\Delta I$ for cases of the "real"\, and "ideal"\, pumps. A correspondent ratio looks like
$$
\frac{\Delta I_{r}}{\Delta I_{i}} = \frac{E_{\pm}}{E_{fl}} \frac{m_{r}}{m_{i}}\sim 10^{4}
$$
it proves once more the advantage of measurement with FSR carrier.\\
For the regime with two component pump such advantage will be increased
drammaticaly. Really modulation indexes for the "real"\, and "duble"\, pump regimes are approximately equal. So the geo-signal magnitude for the "double"\, pump has to be amplified in $(E_{0}/E_{\pm})\sim 10^{6}$ times in respect of "real"\, pump regime. \\  

In discussion of these results one can not avoid the question concerning an influence of the recycling mirror existing in the composition of full interferometer. This scheme also have been analized \cite{16}. So as it requires much more combersome calculation we present here only a brief resultive summary.\\    
Method of consideration was similary to one used in this paper, namely : the full scheme was reduced to some "recycled FP cavity"\, composed by the recycling mirror and equivalent complex mirror of FP-Michelson configuration given at the fig.1. In the geometrical optic approach the transfer function of the "recycled FP cavity"\, depends on the base of cavity $l_{R}$, recycling mirror's reflection and transmission coefficients $r_{R},\tau_{R}$ and similar parameters of the equivalent FP-M mirror $r_{M},\tau_{M}$. The last ones depend on arm's deformations $\xi_{1},\xi_{2}$. The transmission of the equivalent mirror is defined factually by the transfer function of non recycled setup $\tau_{M}\propto G(\omega,\Delta\xi)$ which is equal to zero in the "dark spot"\,regime. However a geophysical information is kept in the reflection parameter $r_{M}=r_{M}(\omega,\xi_{1},\xi_{2})$. Thus the reycled setup allows to extract a geophysical information even in the "dark spot"\, regime:\, a degeneration (compensation) of the liner effect due to the symmetrical influence of neighbouring modes disappears. Adjusting the base parameter $l_{R}$ in coordination with small lengths of the "central michelson interferometer"\, one can make an optimal tuning of the full setup. Under this the intensity of "geophysical signal"\, is increased on the factor of recycled FP cavity finesse. Finally our conclusion concerning - the effect of "geophysical modulation"\, of output harmonics in the spectral zone of GW detection and at the photon circulation frequency - remains to be valied (details see in \cite{16}) .\\

Above we were concentrated at the physical mechanism of penetration low frequency signals at the GW interferometer's output. Let's discuss briefly possible applications to fundamental experiments such as measurement of weak global geophysical effects and detection of low frequency gravitational waves (proposed in \cite{11}).\\
Analysis supposes a comparison of expected signal amplitude with a noise background specific for the given setup. The problem is to get a correct  spectral density of deformational noise background at quasistatic frequencies. Indirect estimate can be extracted from the Virgo geophysical channel (signal of control circuits). From the geo data of VSR2/VSR1 series presented in \cite {8 , 17} after substracting "theoretical tide"\, we have found the estimate of residual strain noise standard on the order of $(10^{-11} - 10^{-12})\, 1/Hz^{1/2}$. Mostly it corresponds to the frequency interval $ (0.1-0.01)\,Hz$
It is well known just in this interval a damping of seismic noise spectral density takes place. As some independent estimate one can use data \cite{18,19} for vertical seismic accelerations in quite areas: $<a_{\omega}>\sim 10^{-7}\, cm/sec^{2}Hz^{1/2}$. Then roughly supposing the similar order of value for horizontal displacements one might estimate its amplitude as $\Delta x\approx (<a_{\omega}> \omega^{-2})\sim(10^{-7}-\,10^{-6})\,cm/Hz^{1/2}$. Deviding it on the interferometer arm length one gets the strain noise at the same level like from the Virgo data. This of course the low limit estimate which we will keep in the mind below analysing the problem of "weak signal detection". \\
a) Geophysical effects. \\
Between weak global geophysical effects which continue to be in the sphere of interest of modern geodynamics there are short tidal harmonics (periods
$(6 - 4)\, h$, high frequency Earth modes (periods $\sim 0.1\,h$) and close group of harmonics composing the "liquid core"\, resonance (periods $\sim 24\,h$) \cite {19}. All this phenomena produce the gravity and strain perturbations at the level $\Delta L/L\sim \Delta g/g \sim \,(10^{-9}-\,10^{-10})$\,($g$\,- the gravity acceleration). In supposition that the strain noise level mentioned above might be extrapolated in the region $(10^{-4}\,- 10^{-5})\,Hz$ one could conclude in favour of a mesurability of these effects in principle. However a real "hot spot"\, of global geodynamics consists in the measurement of the "inner core oscillations" \cite{20}. Period of this fundamental mode $\sim (3 - 4)\,h$ and with the magnitude of displacement from the Globe center $\sim 1\,m $ the expected strain signal would be $\sim (10^{-10}-\,10^{-11})$, i.e. it also might be measured under accumulation data during a few tenths cycle of oscillation.\\
b) Low frequency gravitational waves. \\
Taking into account an existing astrophysical forecast one could not be too optimictic. Frequency interval of our interest $(1\,-\,10^{-4})\,Hz$ partly corresponds to those of LISA project \cite{21}. In the number of GW sources covering this frequency range there are galactic white dwarf (WD) and superdense binaries as well as relativistic stars in the process of inspiral falling down to supermassive black hole in the center of our Galaxy. Statisticaly confidential forecasts for LISA results in the standard GW amplitude less then $h\sim 10^{-20}$ \cite{21}. In attempt to find a more profitable model of source one could consider a very close WD binary (period $\sim 100\,sec$) at the distance 100 pc (forgetting about the "rate of events"): it results in a continious GW radiation with $h\sim 10^{-19}$. Coalescence event of such binary at the Galactic distance $10\, kpc$ produces the GW pulse ($10\,sec.$ duration) also with $h\sim 10^{-19}$ \cite{22}. Only for the enough artificial example: the coalescence NS + BH (ten solar masses) binary at the distance $1\,kpc$ one can wait GW pulse with amplitude $h\sim 10^{-17}$ (the carrier frequency $\sim 1\,Hz$). Even these exagerated sources unlikely might be registered: at the strain background $10^{-12}\,1/Hz^{1/2}$ after one year accumulation of a continious signal one can hope to achieve the sensitivity $h_{min}\approx 2\,10^{-16}1/Hz^{1/2}$ .
However the future (third) generation of GW interferometers placed in a deep underground probably will meet a suppressed seismic and strain noises at least at two order of value \cite{23,24} and a possibility of very low frequency gravitational waves detection on these instruments might be not so hopeless.\\

\textbf{Acknowledgement.}
Authors recognize a stimulating role of prof. A.Mellessinos in the developing this research. We are grateful also to colleagues from EGO-VIRGO team:
prof. A.Giazotto, Jean Yves Vinet and F.Ferrini for multiple discussions.
Our special gratitude belongs to people of La Sapienza group of Virgo:
prof. F.Ricci, S.Frasca and E.Majorana for a help in understanding details
of Virgo setup operation. \\

\begin{center}
\textbf{Appendix A1.\, Method of complex envelope }
\end{center}

For a stationary linear transfering system with a pulse characteristic $g(t)$ an output signal $y(t)$ is coupled with the input one $x(t)$ through the Dugamel
integral relation  
$$
y(t)=\int\limits_{-\infty}^{\infty}g(t-\tau)x(\tau)d\tau\equiv g(t)\ast x(t).
\eqno{(A1)}
$$
Complex envelopes $\tilde{x},\,\tilde{g}(t)$ are defined through equations
$$
x(t)=\mathrm{Re}\left[\tilde{x}(t)\exp\left\{j\omega_{0}t\right\}\right]=
\frac{1}{2}\left[\tilde{x}(t)\exp\left\{j\omega_{0}t\right\}+
\tilde{x}^{*}(t)\exp\left\{-j\omega_{0}t\right\}\right],
$$
$$
g(t)=\mathrm{Re}\left[\tilde{g}(t)\exp\left\{j\omega_{0}t\right\}\right]=
\frac{1}{2}\left[\tilde{g}(t)\exp\left\{j\omega_{0}t\right\}+
\tilde{g}^{*}(t)\exp\left\{-j\omega_{0}t\right\}\right].
$$
Taking into account that $\tilde{g}(t)$ is a slow changing variable comparing with the period $\omega_{0}^{-1}$ one has
$$
\tilde{g}(t)\simeq2g(t)\exp\left\{-j\omega_{0}t\right\}
$$
Instead of the pulse characteristic one can use the transfer function $G(\omega)\leftrightarrow g(t)$ also in a real and complex forms
$$
\tilde{g}(t)\leftrightarrow\tilde{G}(\omega)=2G(\omega_{0}+\omega),\;|\omega|\ll\omega_{0},
$$
above a narrow band character of the transfer function $G(\omega)$ was taken ito account.
It is easy to show that the relation (A1) is valied also for complex envelopes
$$
\tilde{y}(t)=\frac{1}{2}\int\limits_{-\infty}^{\infty}\tilde{g}(t-\tau)\tilde{x}
(\tau)d\tau=\frac{1}{2}\tilde{g}(t)\ast\tilde{x}(t).
$$

\begin{center}
\textbf{Appendix A2. \,Interferometer output field }
\end{center}
The pump at the OS input is 
$$
E(t)=E_{0}\cos\left[\omega_{0}t+\beta\sin\bar{\Omega}t\right]=
\mathrm{Re}\left[\tilde{E}(t)\exp\left\{j\omega_{0}t\right\}\right],
$$
$$
\tilde{E}=E_{0}\exp\left\{j\beta\sin\bar{\Omega}t\right\}.
$$
Using the well konwn expansion
$$
\exp\left\{jz\sin t\right\}=\sum_{m=-\infty}^{\infty}J_{m}(z)\exp\left\{jmt\right\},\;
J_{-m}(z)=(-1)^{m}J_{m}(z).
$$
one comes to
$$
\tilde{E}\propto J_{0}(\beta)+J_{1}(\beta)\left[\exp\left\{j\bar{\Omega}t\right\}-
\exp\left\{-j\bar{\Omega}t\right\}\right]+\ldots.
$$
Thus the coupling of output and input fields looks like
$$
E_{2}=\tilde{E}_{1}(t)\ast\tilde{g}(t)\propto J_{0}(\beta)\bar{G}(0)+J_{1}(\beta)\left[\tilde{G}(\bar{\Omega})\exp\left\{j\bar{\Omega}t\right\}-
\tilde{G}(-\bar{\Omega})\exp\left\{-j\bar{\Omega}t\right\}\right].
$$
Replacing the exponents by harmonical functions \\  
$(\exp\left\{\pm j\bar{\Omega}t\right\}=\cos\bar{\Omega}t\pm j\sin\bar{\Omega}t)$ leads to the formula (4) in the main text.

\begin{figure}
\begin{center}
\includegraphics[width=8cm]{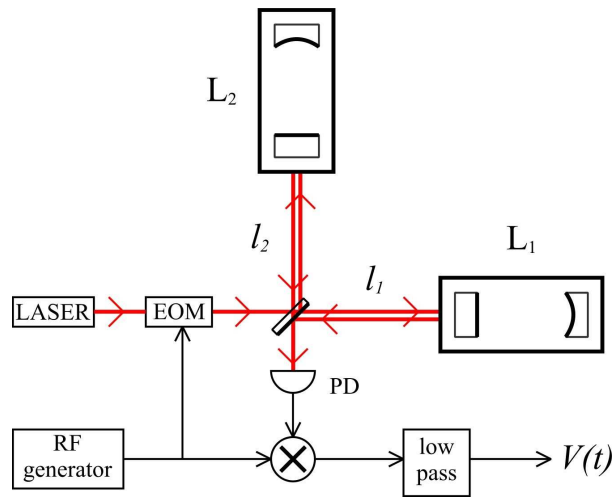}
\caption{Layout of a gravitational wave interferometer: $l_1$, $l_2$ - small michelson arms; $L_1$, $L_2$ - equivalent FP-cavity arm mirrors}
\end{center}
\end{figure}

\end{document}